\documentclass[10pt]{iopart}
\usepackage{iopams,amsmath2,bbm}
\usepackage{graphicx,color,listings}

\newcommand{\tn}{\textnormal}
\newcommand{\cprb}[3]{Phys.~Rev.~B {\bf #1}, #2 (#3)}
\newcommand{\cprl}[3]{Phys.~Rev.~Lett.~{\bf #1}, #2 (#3)}

\newcommand{\cpre}[3]{Phys.~Rev.~E {\bf #1}, #2 (#3)}

\definecolor{darkred}{rgb}{0.90,0,0}
\definecolor{darkgreen}{rgb}{0,0.60,.2}
\definecolor{darkblue}{rgb}{0,0,1}
\definecolor{pink}{rgb}{1,0,1}
\definecolor{grey}{cmyk}{0,0,0,0.25}
\definecolor{orange}{cmyk}{0,0.6,0.8,0}

\begin{document}
\title[Extending the range of real time DMRG simulations]{Extending the range of real time density matrix renormalization group simulations}

\author{D.\ M.\ Kennes$^1$ and C.\ Karrasch$^{2,3}$}

\address{$^1$Institut f\"ur Theorie der Statistischen Physik, RWTH Aachen University and JARA---Fundamentals of Future Information Technology, 52056 Aachen, Germany}

\address{$^2$Department of Physics, University of California, Berkeley, CA 95720, USA}

\address{$^3$Materials Sciences Division, Lawrence Berkeley National Laboratory, Berkeley, CA 94720, USA}

\begin{abstract}
We discuss a few simple modifications to time-dependent density matrix renormalization group (DMRG) algorithms which allow to access larger time scales. We specifically aim at beginners and present practical aspects of how to implement these modifications almost effortlessly within any standard matrix product state (MPS) based formulation of the method. Most importantly, we show how to `combine' the Schr\"odinger and Heisenberg time evolutions of arbitrary pure states $|\psi\rangle$ and operators $A$ in the evaluation of $\langle A\rangle_\psi(t)=\langle\psi|A(t)|\psi\rangle$. This includes quantum quenches. The generalization (non-)thermal mixed state dynamics $\langle A\rangle_\rho(t)=\tn{Tr}[\rho A(t)]$ induced by an initial density matrix $\rho$ is straightforward. In the context of equilibrium (ground state or finite temperature $T>0$) correlation functions, one can extend the simulation time by a factor of two by `exploiting time translation invariance', which is efficiently implementable within MPS DMRG. We present a simple analytic argument for why a recently-introduced disentangler succeeds in reducing the effort of time-dependent simulations at $T>0$. Finally, we advocate the python programming language as an elegant option for beginners to set up a DMRG code.
\end{abstract}

\maketitle


\section{Introduction}
\label{sec:intro}

The density matrix renormalization group (DMRG) \cite{dmrgrev0,dmrgrev} was originally devised \cite{white1,white2} as a tool to accurately determine \textit{static} ground state properties of one dimensional systems. From a modern perspective, the core DMRG algorithm can be formulated elegantly as a variational calculation of the ground state $|\tn{gs}\rangle$ within the class of matrix product states (MPS) \cite{mps1,mps2,verstraete1,verstraete2},
\begin{equation}\label{eq:mps}
|\psi\rangle = \sum_{\{\sigma_{l}\}} M^{\sigma_1}\cdot M^{\sigma_2} \cdots
M^{\sigma_L} |\{\sigma_{l}\}\rangle \,.
\end{equation}
Due to the area law \cite{arealaw,arealaw2}, a fairly small matrix (`bond') dimension $\chi$, which encodes the amount of entanglement, is sufficient to describe $|\tn{gs}\rangle$ accurately. One way to obtain \textit{correlation functions},
\begin{equation}\label{eq:corrgs}
C^{AB}_\tn{gs}(t)=\langle\tn{gs}|A(t)B|\tn{gs}\rangle\,,~A(t) = e^{iHt}Ae^{-iHt}\,,
\end{equation}
or to simulate \textit{non-equilibrium (quench) dynamics}
\begin{equation}\label{eq:quench}
\langle A\rangle_\psi(t) = \langle\psi|A(t)|\psi\rangle
\end{equation}
in an arbitrary state $|\psi\rangle$, is to directly calculate the real time evolution using a time-dependent DMRG framework \cite{tdmrg1,tdmrg2,tdmrg3,tdmrg3a,tdmrg4,tebd0,tebd}\footnote{For other approaches see Refs.~\cite{dmrgrev0,dyndmrg0,dyndmrg0a,dyndmrg,jeckelmann,jan,piet,haegeman}.}. The corresponding algorithm can again be formulated elegantly using matrix product states. The physical growth of entanglement implies that the bond dimension needed to approximate $C^{AB}_\tn{gs}(t)$ or $\langle A\rangle_\psi(t)$ to a certain accuracy grows with time, often even exponentially. This limits the accessible time scales. Modified algorithms such as transverse folding \cite{verstraete0} hold the promise of substantially extending the range of simulations, but implementing them in practice requires some effort (for a comprehensive overview of other available approaches see Ref.~\cite{dmrgrev} and references therein). It is the main goal of this paper to discuss a few \textit{simple} `recipes' that allow to reach larger times in DMRG calculations. We specifically aim at an audience of beginners and ask: Assuming that one has a standard MPS based DMRG code at hand, what are the most important practical steps necessary to incorporate these recipes? For colleagues new to the realm of DMRG, we try to advocate the method in general by showing how straightforwardly its core algorithms can be implemented within the python programming language.

\textit{A factor of two} ---
The `standard way' to obtain the real time correlation function $C^{AB}_\tn{gs}(t)$ is to compute $e^{-iHt}B|\tn{gs}\rangle$. However, one can simply `exploit time translation invariance' in this \textit{equilibrium} problem, recast Eq.~(\ref{eq:corrgs}) as
\begin{equation}\label{eq:trickC}
C_\tn{gs}^{AB}(2t) = \langle\tn{gs}| A(t) B(-t) | \tn{gs}\rangle\,,
\end{equation}
and carry out two separate DMRG simulations for $e^{-iHt}B|\tn{gs}\rangle$ and $e^{iHt}A|\tn{gs}\rangle$, respectively. This allows to reach time scales twice as large as before at no additional effort. To the best of our knowledge, this was overlooked for a long time and only noticed implicitly in a recent paper \cite{trick2b} in the specific context of optimizing calculations at finite temperature $T$ (see below). We will now explain practical aspects of how a similar `trick' can be implemented in \textit{non-equilibrium} to push the quench dynamics described by Eq.~(\ref{eq:quench}) to larger times.

\textit{Schr\"odinger vs.~Heisenberg picture ---}
The most straightforward way to evaluate Eq.~(\ref{eq:quench}) within DMRG is to simulate $e^{-iHt}|\psi\rangle$. This corresponds to a time evolution in the Schr\"odinger picture. If one generalizes the concept of a matrix product state to a matrix product operator (MPO), one can alternatively switch to the Heisenberg picture and calculate the operator time evolution $e^{iHt}Ae^{-iHt}$ \cite{mpodmrg1,mpodmrg2,mpodmrg3,mpodmrg3a,mpodmrg4,mpodmrg5,mpodmrg6,mpodmrg7}. This is equivalent mathematically but different algorithmically since $e^{iHt}Ae^{-iHt}$ might have a more efficient representation in terms of a MPO than $e^{-iHt}|\psi\rangle$ has in terms of a MPS (or vice versa). Indeed, it was shown \cite{mpodmrg2} that the time evolution of $S^z_l$ under the spin-$1/2$ XXZ Hamiltonian
\begin{equation}\label{eq:h}
H = \sum_{l=1}^{L-1} \left[\frac{1}{2}\left(S^+_lS^-_{l+1} + S^-_lS^+_{l+1}\right) + \Delta S^z_lS^z_{l+1} \right] + b \sum_{l=1}^{L} S^z_l
\end{equation}
can be expressed exactly in terms of a MPO with a finite bond dimension if $\Delta=0$. Thus, $\langle A=S^z_l\rangle_\psi(t)$ can be simulated up to arbitrary long times for any state $|\psi\rangle$ using the Heisenberg picture, and certainly also for any $A$ and $\Delta\neq0$ if $|\psi\rangle$ is an eigenstate of $H$ using the Schr\"odinger picture. For a general scenario in between those two extreme limits, however, one would expect that the bond dimension $\chi$ increases equally fast during the calculation of $e^{-iHt}|\psi\rangle$ and $e^{iHt}Ae^{-iHt}$. In this case one can simply split the time evolution between the Schr\"odinger- and Heisenberg picture,
\begin{equation}\label{eq:trickA}
\langle A\rangle_\psi(2t) = \langle\psi| e^{iHt} A(t) e^{-iHt} |\psi\rangle\,, 
\end{equation}
and evaluate $e^{-iHt}|\psi\rangle$ as well as $A(t)$ individually. We will present a few generic examples below (including the more general situation of quenches at finite temperatures) and show that in many physical applications Eq.~(\ref{eq:trickA}) actually allows to access time scales approximately twice as large as before \textit{using the same bond dimension}. In light of the fact that $\chi$ often grows exponentially fast, this amounts to major algorithmic savings. As a guide for beginners, we will point out the most important practical aspects of how to implement the calculation of $e^{iHt} A e^{-iHt}$ within an existing MPS based DMRG code and discuss the XXZ spin chain as well the Hubbard model of interacting lattice fermions $c_{l\sigma}$,
\begin{equation}\label{eq:h2}\begin{split}
H = \sum_{l}\Big\{& -\sum_\sigma \Big[\frac{1}{2}  c_{l\sigma}^\dagger c_{l+1\sigma}^{\phantom{\dagger}} + \tn{h.c.} \Big]
+ Un_{l\uparrow}n_{l\downarrow} \\
& +\mu \big( n_{l\uparrow}+n_{l\downarrow}\big)
+b \big( n_{l\uparrow}-n_{l\downarrow}\big)\Big\}\,,~~
n_{l\sigma}= c_{l\sigma}^\dagger c_{l\sigma}^{\phantom{\dagger}}-\frac{1}{2}\,,
\end{split}\end{equation}
as two prototypical examples. We will particularly elaborate how to incorporate all Abelian symmetries \cite{mpodmrg3a} \footnote{As a side product, we will demonstrate how to directly exploit multiple Abelian symmetries (e.g., spin and charge conservation for the Hubbard model) within an existing time-dependent DMRG code without having to modify it at all.}. This altogether provides a simple recipe to potentially extend the range of simulations at virtually no (or very little) effort.

\textit{Finite temperatures ---}
Standard DMRG methods allow computing the time evolution of a \textit{pure} state and are thus not directly applicable at finite temperatures. In order to simulate dynamics at $T>0$, one can use operator space DMRG \cite{trick2b,trick2a}, or -- equivalent mathematically -- one can express the thermal statistical operator $\rho_T\sim e^{-H/T}$ as a partial trace over a pure state $|\Psi_T\rangle$ living in an enlarged Hilbert space where auxiliary degrees of freedom $Q$ encode the thermal bath \cite{purification,dmrgT,barthel,fiete,drudepaper}:\footnote{Various other ways to incorporate finite temperatures within DMRG can be found in Refs.~\cite{verstraete,vidalop,tmrg0,tmrg0a,metts,tmrg1,sirker1}.}
\begin{equation}
\rho_T = \tn{Tr}_Q |\Psi_T\rangle\langle\Psi_T|\,. 
\end{equation}
A finite-$T$ correlation function can in principle be obtained straightforwardly by carrying out real- and imaginary time evolutions of the (known) state $|\psi_\infty\rangle$ which purifies $\rho_T$ at $T=\infty$ \cite{dmrgrev}: 
\begin{equation}\label{eq:corrt}
C^{AB}_T(t) = \tn{Tr}\,\big[\rho_T A(t)B\big] =
\langle\psi_T|A(t)B |\psi_T\rangle \sim
\langle\psi_\infty|e^{-H/2T}A(t)Be^{-H/2T} |\psi_\infty\rangle\,.
\end{equation}
In practice, however, the time scales accessible at nonzero $T$ are considerably smaller than those at $T=0$ \cite{barthel}. In Ref.~\cite{drudepaper} it was shown that one can exploit the fact that purification is not unique to push simulations to larger times.\footnote{An alternative approach to potentially extend the range of simulations substantially is the probabilistic sampling over an appropriately chosen set of pure states introduced in Ref.~\cite{metts}.} In particular, one can insert any unitary transformation $U_Q(t):Q\to Q$ which solely acts on the auxiliary Hilbert space $Q$ into Eq.~(\ref{eq:corrt}):
\begin{equation}\label{eq:aux}\begin{split}
\rho_T =  U_Q^{\phantom{\dagger}}(t) U_Q^\dagger(t) \rho_T = U_Q^{\phantom{\dagger}}(t) \rho_T U_Q^\dagger(t) & = \tn{Tr}_Q U_Q^{\phantom{\dagger}}(t) |\Psi_T\rangle\langle\Psi_T|U_Q^\dagger(t)\,, \\
\Rightarrow~ C^{AB}_T(t) & = \langle\psi_T|U_Q^\dagger(t) A(t)U_Q^{\phantom{\dagger}}(t)B |\psi_T\rangle\,.
\end{split}\end{equation}
It turned out that choosing $U_Q(t) = e^{iH_Qt}$, i.e. a time evolution in $Q$ governed by the physical Hamiltonian (where physical degrees of freedom are replaced by auxiliary ones) but reversed time leads to a slower increase of the bond dimension, and thus longer time scales can be reached (a systematic way to further optimize $U_Q(t)$ was introduced in Refs.~\cite{trick2b,trick2a}; particularly at low temperatures, this allows to access larger times). In the realm of Ref.~\cite{drudepaper}, however, the specific form of $U_Q(t) = e^{iH_Qt}$ was nothing but a `lucky guess'. Eq.~(\ref{eq:aux}) was subsequently mapped to a problem of time-evolving operators \cite{trick2b}, which then provided an understanding of the `disentangler' $U_Q(t)$. While mathematically equivalent, we aim at reformulating the argument of Ref.~\cite{trick2b} in the \textit{simplest possible} way for didactic purposes. More importantly, this yields an explanation for why the signs of certain terms in $H_Q$ need to be reversed (and provides a strategy to determine which ones) if symmetries are exploited in the MPS formalism.

The finite-temperature analogue of Eq.~(\ref{eq:trickC}) reads \cite{trick2b,trick2a}
\begin{equation}\label{eq:trickCT}
C^{AB}_T(2t) = \tn{Tr}\,\big[\rho_T A(t)B(-t)\big] =
\langle\psi_T|A(t)B(-t) |\psi_T\rangle\,.
\end{equation}
In principle, $\langle\psi_T|A(t)B(-t) |\psi_T\rangle$ can be computed straightforwardly within any MPS based DMRG formalism. In practice, however, the forward and backward time evolutions in $e^{iHt_n}Ae^{-iHt_n}|\psi_T\rangle$ need to be carried out individually for each $t_n$ in order to obtain $\langle\psi_T|A(t_n)B(-t_n) |\psi_T\rangle$ for all $t_n\in[0,t]$ -- a problem which does not occur for the ground state correlation function of Eq.~(\ref{eq:trickC}). Moreover, if $A$ can only be written as a sum of $m>1$ local operators, one generally needs to evaluate each term separately. We will illustrate below that both problems can be circumvented and that Eq.~(\ref{eq:trickCT}) can be implemented effortlessly within MPS based DMRG.

In the remainder of this paper we elaborate on the above-mentioned issues and present concrete examples. We perform DMRG calculations using a fixed small discarded weight and a 4th order Trotter decomposition \cite{suzuki} of the time evolution operators. As our paper aims at an audience familiar with basics of the method, we refrain from giving a more detailed introduction but refer the reader to Refs.~\cite{dmrgrev,dmrgpaper}.

\begin{figure}[t]
\includegraphics[width=0.48\linewidth,clip]{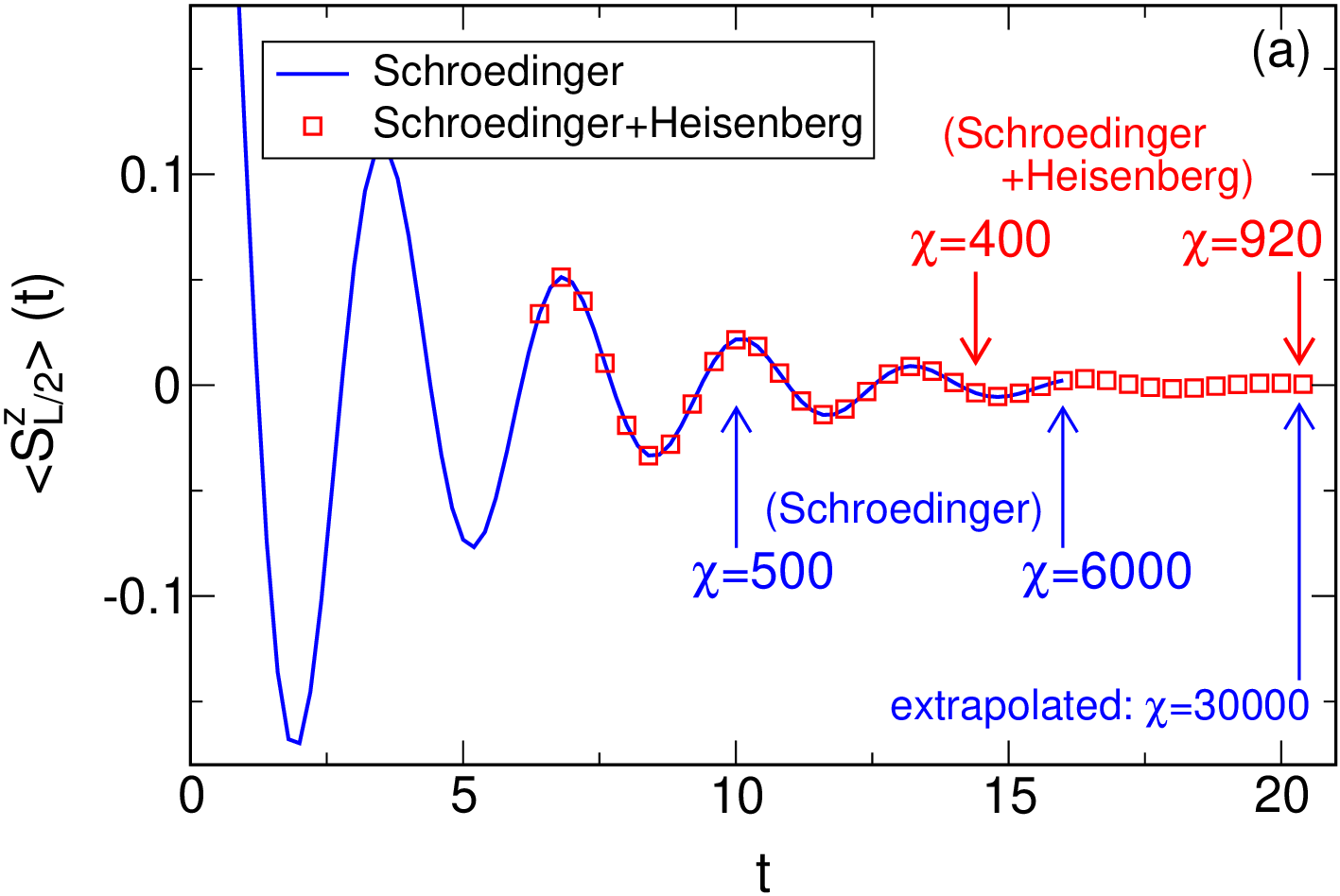}\hspace*{0.035\linewidth}
\includegraphics[width=0.48\linewidth,clip]{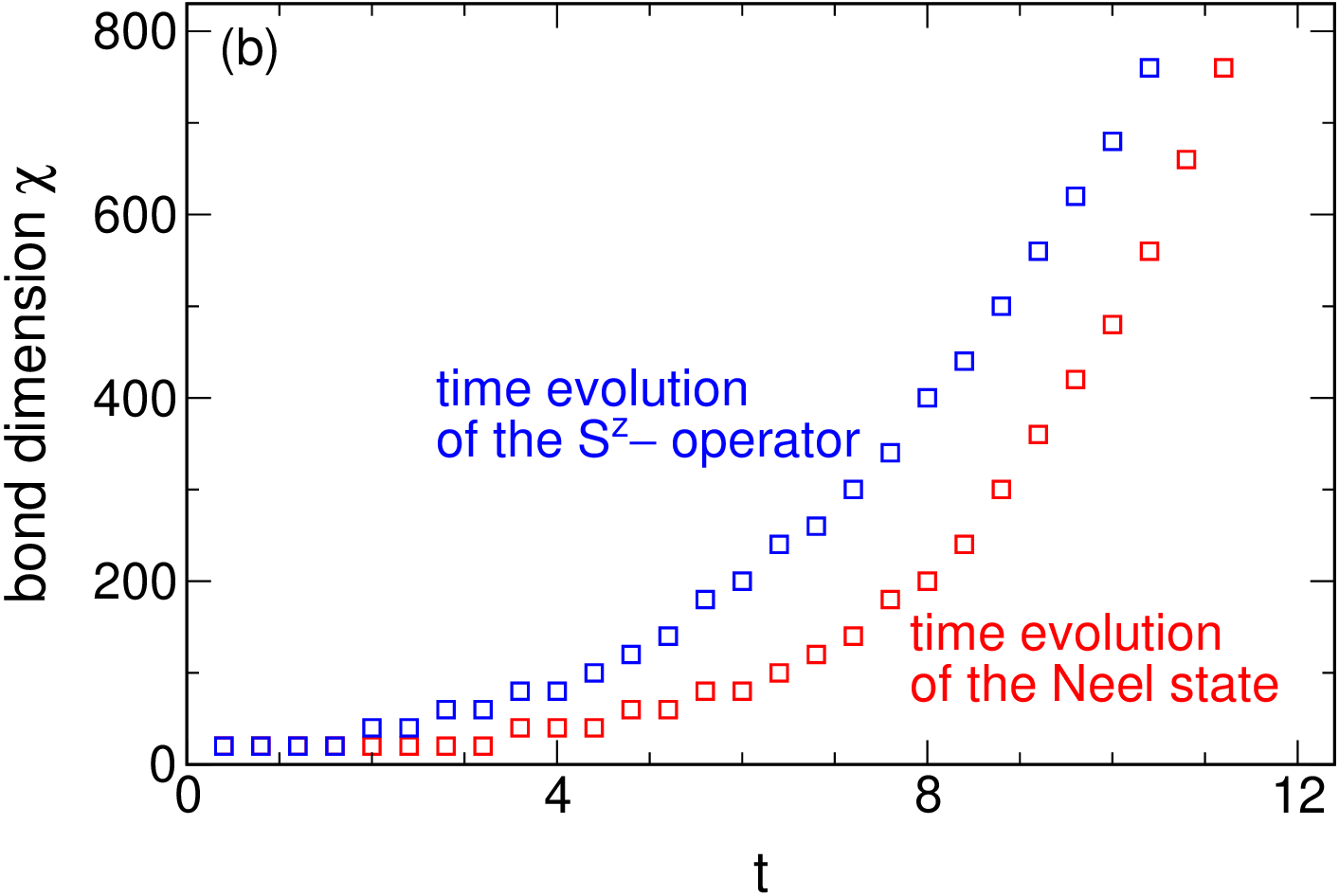}
\caption{(Color online) (a) DMRG calculation of the spin expectation value $\langle S^z_{L/2}\rangle_\tn{NS}(t)= \langle \tn{NS}|S^z_{L/2}(t)|\tn{NS}\rangle$ in a Neel state $|\tn{NS}\rangle$ under the XXZ Hamiltonian of Eq.~(\ref{eq:h}) with $\Delta=0.5$, $b=0$, and $L=100$. If the time evolution is split equally between the Schr\"odinger and Heisenberg pictures via Eq.~(\ref{eq:trickA}), the simulation can be pushed to larger times with no additional effort. (b) Growth of the bond dimension $\chi$ during the evaluation of $e^{-iHt}|\tn{NS}\rangle$ and $S^z_{L/2}(t)$. Note that the latter can be computed straightforwardly within any existing MPS based implementation of the method (see Sec.~\ref{sec:mpo} for details).}
\label{fig:neel}
\end{figure}

\section{Schr\"odinger vs.~Heisenberg picture}
\label{sec:heisenberg}

We study the time evolution of an observable $\langle A\rangle_\psi(t)= \langle \psi| e^{iHt}Ae^{-iHt}|\psi\rangle$ in an arbitrary state $|\psi\rangle$ as well as the more general scenario of a quench at finite temperature,
\begin{equation}
\langle A\rangle_\rho(t)=\tn{Tr}\left[\rho e^{iHt}Ae^{-iHt}\right]\,,
\end{equation}
where $\rho$ is some initial density matrix. We focus on the XXZ spin chain defined by Eq.~(\ref{eq:h}), which can be mapped to spinless lattice fermions with a nearest-neighbor Coulomb interaction $\Delta$. If one `combines' the Schr\"odinger- and Heisenberg pictures through Eq.~(\ref{eq:trickA}) or its $T>0$ analogue,
\begin{equation}\label{eq:trickAT}
\langle A\rangle_\rho(2t) =  \tn{Tr}\left[\rho(-t)A(t)\right]\,,
\end{equation}
calculates $e^{-iHt}|\psi\rangle$ or $\rho(-t)$ as well as $A(t)$ separately, and stops each simulation at times $t_{\psi/\rho}$ and $t_A$ where a fixed bond dimension is reached, one can \textit{always} reach larger time scales $t_{\psi/\rho}+t_A>t_{\psi/\rho},t_A$.\footnote{This is not hindered by a more costly `overlap' calculation; see Sec.~\ref{sec:mpo} for details.} In practice, however, $t_{\psi/\rho}$ might be significantly smaller than $t_A$ or vice versa, and the additional effort to implement Eqs.~(\ref{eq:trickA}) and (\ref{eq:trickAT}) might not be justified. This is certainly the case if either $|\psi\rangle$ is close to an eigenstate of $H$ (if $\rho$ is close to the thermal density matrix), or if the time evolution of $A$ can be expressed efficiently in terms of a MPO (e.g., for $A=S^z_l$ and $\Delta=0$). However, one would expect that for a large class of problems both simulations are `equally complex', and that thus $t_{\psi/\rho}\approx t_A$.  We will now study this for two generic physical problems.

We first investigate the time evolution of $S^z_l$ in a Neel state $|\tn{NS}\rangle$, which is an example for a \textit{global quantum quench}. We know that $\langle S^z_l\rangle_\tn{NS}$ can be computed trivially 1) in the Schr\"odinger picture for $\Delta=\infty$, or 2) in the Heisenberg picture for $\Delta=0$. For a generic value of $\Delta=O(1)$, however, the bond dimension grows equally fast during the evaluation of $e^{-iHt}|\tn{NS}\rangle$ and $S^z_l(t)$. This is shown in Figure \ref{fig:neel}(a,b) for $\Delta=0.5$ (results for $\Delta=0.25$ and $\Delta=1$ look similar). Combining the Schr\"odinger and Heisenberg picture thus allows to access time scales roughly twice as large [see Figure \ref{fig:neel}(a)] using the same $\chi$. Since $\chi$ typically grows exponentially fast (by extrapolation one can estimate the bond dimension required to reach $t\sim20$ within the Schr\"odinger picture to be $\chi\sim30000$), this amounts to major algorithmic savings.

\begin{figure}[t]
\includegraphics[width=0.48\linewidth,clip]{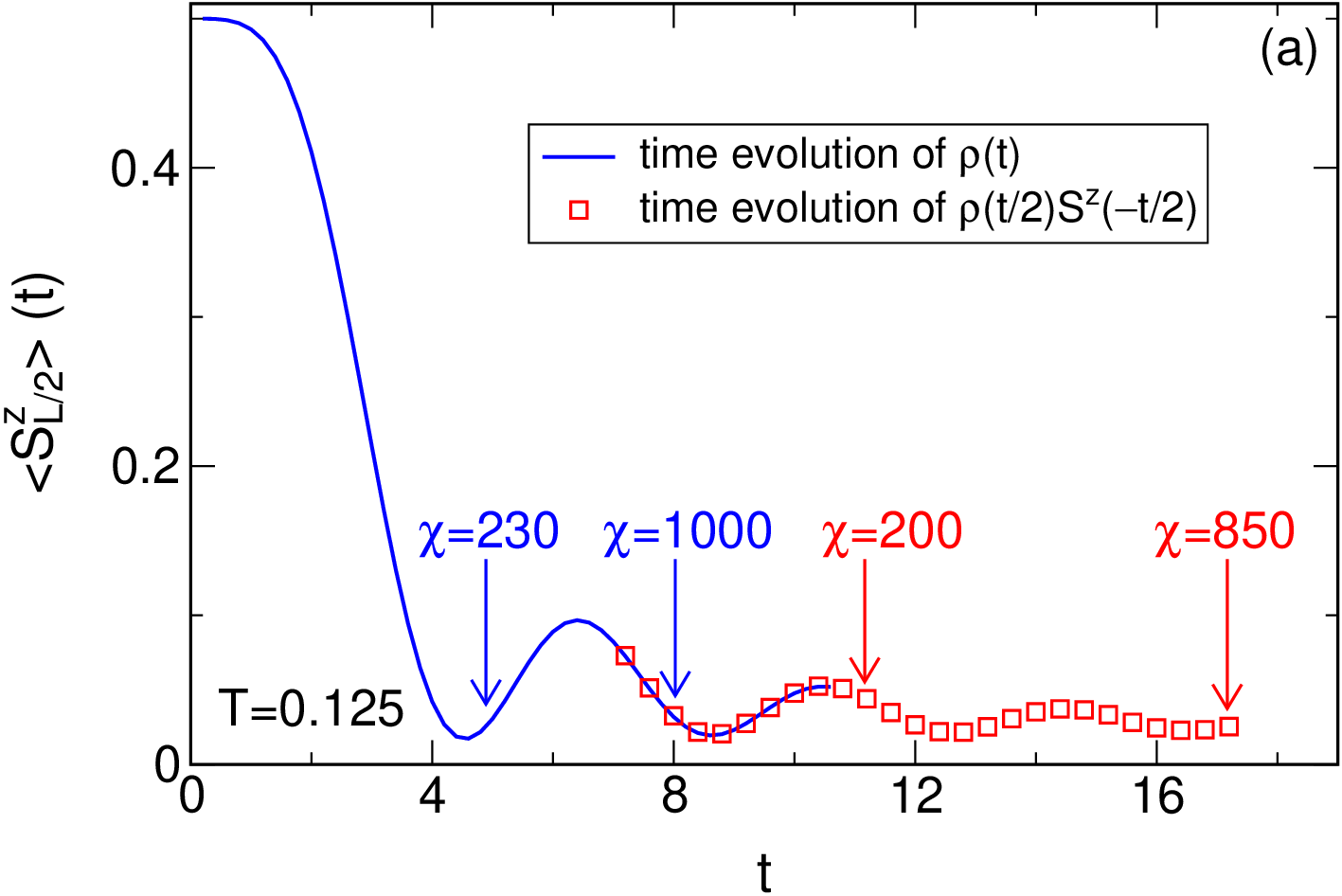}\hspace*{0.035\linewidth}
\includegraphics[width=0.48\linewidth,clip]{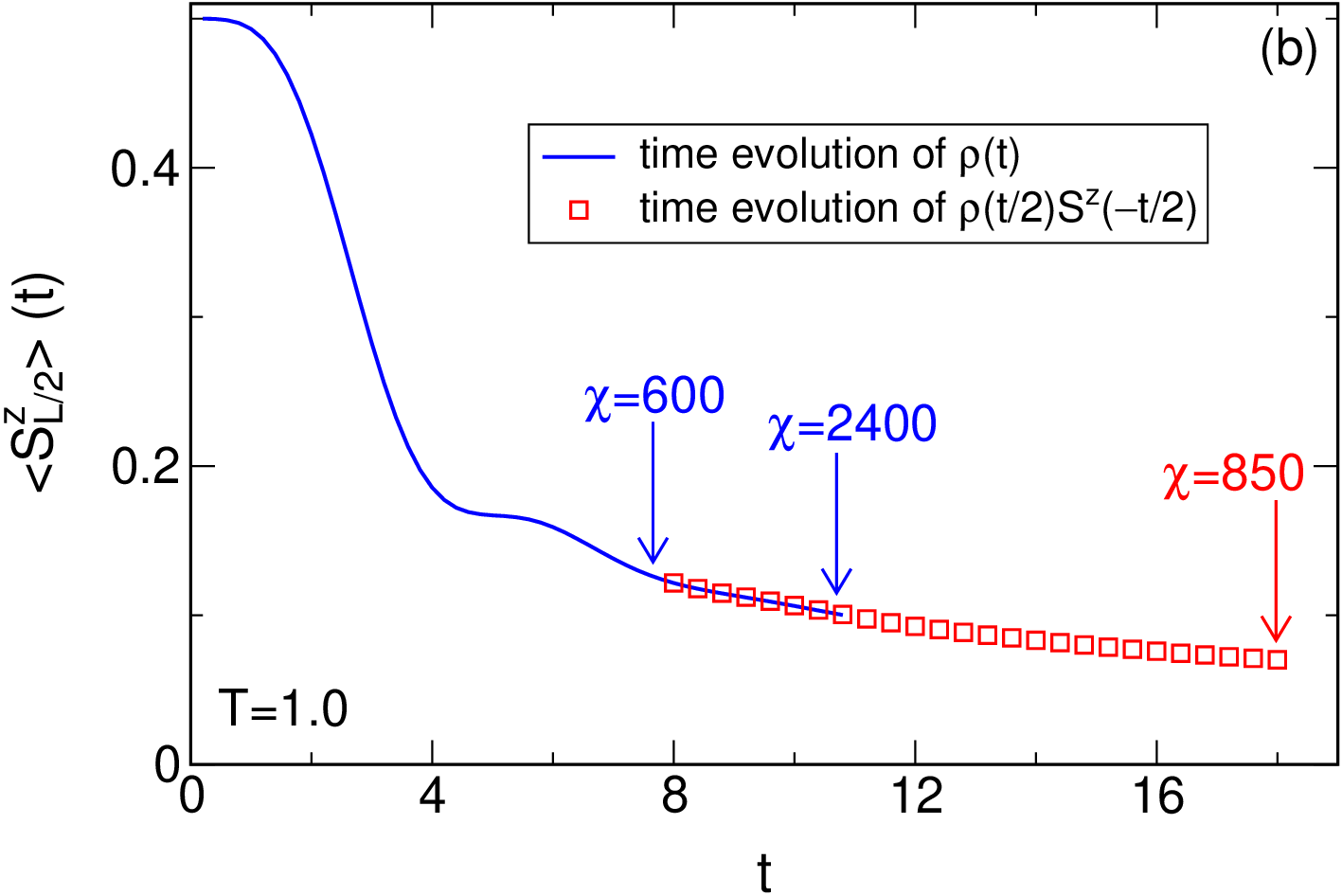}
\caption{(Color online) Spin expectation value $\langle S^z_{L/2}\rangle_\rho(t)$ following a local quench at finite temperatures. At time $t=0$, a `wave packet' of four up-spins is prepared at the center of an isotropic XXZ chain ($\Delta=1$, $b=0$) of length $L=100$ which is otherwise in thermal equilibrium. This initial geometry is described by a density matrix $\rho=\rho^L_T\otimes\rho_{\uparrow\uparrow\uparrow\uparrow}\otimes\rho^R_T$. One can calculate $\langle S^z_{l}\rangle_\rho(t)=\tn{Tr}[\rho e^{iHt} S^z_{l}e^{-iHt}]$  either by a straightforward time evolution of the state which purifies $\rho$ \cite{finiteTquench}, or using $\langle S^z_{l}\rangle_\rho(t)=\tn{Tr}[\rho(-t/2) S^z_{l}(t/2)]$ and time-evolving $\rho(-t/2)$ as well as $S^z_{l}(t/2)$ separately. The latter allows to perform the simulation up to larger times. }
\label{fig:wavep}
\end{figure}

As a second example, we study a \textit{local quench} at finite temperature described by an initial density matrix $\rho=\rho_T^L\otimes\rho_{\uparrow\uparrow\uparrow\uparrow}\otimes\rho_T^R$. This models a wave packet of four up-spins surrounded to its left and right by chains in thermal equilibrium. We focus on the isotropic XXZ chain ($\Delta=1$) and compute the time evolution of $S^z_l$ via
\begin{equation}
\langle S^z_l\rangle_\rho(t) = \tn{Tr}\left[\rho e^{iHt} S^z_le^{-iHt} \right] = \langle \psi_\rho| e^{iHt} S^z_le^{-iHt} |\psi_\rho\rangle\,, ~~\rho=\tn{Tr}_Q|\psi_\rho\rangle\langle\psi_\rho|\,,
\end{equation}
where the state $|\psi_\rho\rangle$ which purifies $\rho$ can be obtained straightforwardly from $|\psi_T\rangle$ and the trivial purification of $\rho_{\uparrow\uparrow\uparrow\uparrow}$. We separately calculate $e^{-iHt}e^{iH_Qt}|\psi_\rho\rangle$, where we insert $e^{iH_Qt}$ to reduce the buildup of entanglement \cite{drudepaper,dmrgpaper}, as well as $S^z_l(t)$. As illustrated in Figure \ref{fig:wavep}, the bond dimension grows comparably fast during both simulations, and thus larger time scales can be reached if they are combined. Note that the `standard' approach (e.g., used in Ref.~\cite{finiteTquench}) is to time-evolve only the state $|\psi_\rho\rangle$ but not $S^z_l$.

\section{Operator time evolution within MPS DMRG}
\label{sec:mpo}

In this section, which solely aims at DMRG beginners, we discuss a few practical aspects of how to simulate the operator time evolution $A(t)$ \cite{mpodmrg1,mpodmrg2,mpodmrg3,mpodmrg3a,mpodmrg4,mpodmrg5,mpodmrg6,mpodmrg7} using an existing MPS based DMRG implementation. Put differently: We want to provide a simple practical guide of how to calculate $A(t)$ assuming that one has a standard DMRG code at hand which allows to compute the time evolution of a state. We start by noting that any $A$ can (in principle) always be expressed as a MPO
\begin{equation}\label{eq:mpo}
A = \prod_{l=1}^{L} A^{[l]} = \sum_{\{\sigma_{l}\}}\sum_{\{\tilde\sigma_{l}\}} A^{\sigma_1,\tilde\sigma_1}\cdot A^{\sigma_2,\tilde\sigma_2} \cdots
A^{\sigma_L,\tilde\sigma_L} |\{\sigma_{l}\}\rangle\langle\{\tilde\sigma_{l}\}|\,
\end{equation}
if one allows for an exponentially large bond dimension $\chi\sim e^L$ \cite{dmrgrev}. Fortunately, most physical observables have a representation in terms of a MPO with a small $\chi=O(1)$ that can be obtained by mere inspection; e.g.~$S^z_l = \mathbbm{1}\cdots \mathbbm{1}\cdot S^z_l\cdot \mathbbm{1} \cdots \mathbbm{1}$ is a trivial MPO with $\chi=1$ (see Sec.~\ref{sec:fac2} for a more complex example). We will now discuss two alternative approaches of how to simulate $A(t)$ provided that the MPO representation of $A$ is known. In particular, we will illustrate that Abelian symmetries can be exploited straightforwardly; it is thus instructive to recapitulate how such symmetries are generally incorporated within DMRG numerics (we follow Ref.~\cite{dmrgrev}). To this end, let us consider an eigenstate $|\psi\rangle$ of an operator $M=\sum M_l$ which defines an additive local quantum number, $M_l|\sigma_l\rangle = m(\sigma_l)|\sigma_l\rangle$. One can show recursively that $|\psi\rangle$ can be expressed as a MPS whose `block states'
\begin{equation}\label{eq:block}
|a_{l}\rangle = \sum_{\{\sigma_{i \leq l}\}} (M^{\sigma_1}\cdots M^{\sigma_{i}})_{1,a_{l}} |\{\sigma_{i}\}\rangle \,
\end{equation}
are eigenstates of $\sum_{l\leq l_0}M_l$ with a quantum number of $m^\Sigma_{l}$; this implies that all matrix elements in Eq.~(\ref{eq:mps}) vanish except for those with
\begin{equation}\label{eq:sym}
m^\Sigma_{l} + m(\sigma_{l+1}) = m^\Sigma_{l+1}\,,
\end{equation}
and the same obviously holds for the time-evolved state $e^{-iHt}|\psi\rangle$ if $[H,M]=0$. Eq.~(\ref{eq:sym}) can be readily incorporated within a DMRG code\footnote{While exploiting Abelian symmetries within the MPS numerics is straightforward, incorporating continuous non-Abelian symmetries is much more involved \cite{sym1,sym2,sym3}.} to significantly reduce the computational effort -- for the problems studied in this work, numerics speed up by a factor of 10 for a bond dimension of $\chi\sim1000$. If the model at hand features more than one Abelian symmetry, this can be exploited within any existing code without having to modify it at all: In case of the Hubbard model where both the total spin and charge are conserved, one simply assigns the following quantum numbers to the states $\{|0\rangle,|\hspace*{-0.1cm}\uparrow\rangle,|\hspace*{-0.1cm}\downarrow\rangle,|\hspace*{-0.1cm}\uparrow\downarrow\rangle\}$ which span its local Hilbert space:
\begin{equation}
m(0) = 0\,,~m(\uparrow) = 1 \,,~m(\downarrow) = M_0\,,~m(\uparrow\downarrow) = M_0+1~,
\end{equation}
where $M_0>2L$ is an arbitrary integer. This automatically accounts for both spin and charge conservation ($M_0>2L$ guarantees the separation of the two symmetries). Other models/symmetries can be treated analogously.

\textit{The trivial way} --- The coefficients $A^{\sigma_1,\tilde\sigma_1} \cdots A^{\sigma_L,\tilde\sigma_L}$ appearing in Eq.~(\ref{eq:mpo}) are nothing but the coefficients of a matrix product state whose local Hilbert space is $d^2$-dimensional and parametrized by a superindex $\Sigma_l=(\sigma_l\tilde\sigma_l)$. The time evolution operators $e^{iHt}$ (acting on $\sigma_l$) as well as $e^{-iHt}$ (acting on $\tilde\sigma_l$) can thus be applied directly using an existing MPS based DMRG code \cite{mpodmrg3a}. The only (minor) subtlety is how to incorporate Abelian symmetries (in case that the initial $A$ respects them) to speed up the calculation. If $M=\sum_lM_l$ is conserved , this can be exploited by assigning a quantum number $m(\sigma\tilde\sigma)=m(\sigma)-m(\tilde\sigma)$ to the local states $\{|\sigma\tilde\sigma\rangle\}$ within the MPS numerics; for a spin-$1/2$ system ($d=2$) and $M=S^z$, the $m$'s read
\begin{equation}
m(\Sigma=\uparrow\uparrow)= 0\,,~ m(\Sigma=\downarrow\downarrow)= 0\,,~m(\Sigma=\uparrow\downarrow)= 1\,,~m(\Sigma=\downarrow\uparrow)= -1\,.
\end{equation}
Other symmetries follow analogously. The computational effort of this approach scales as $d^6\chi^3$ \cite{dmrgrev}.

\textit{Employing finite-$T$ numerics} --- For models with a large local Hilbert space dimension (such as the Hubbard model where $d=4$) one can resort to an alternative approach. After recasting the coefficients in Eq.~(\ref{eq:mpo}) via a singular value decomposition (SVD),
\begin{equation}\begin{split}
[A^{\sigma_l,\tilde\sigma_l}]_{a_l,a_{l+1}} = A_{(a_l\sigma_l),(a_{l+1}\tilde\sigma_l)} & \stackrel{\tn{SVD}}{=}
\sum_{s_l} U_{(a_l\sigma_l),s_l} S_{s_l} V_{s_l,(a_{l+1}\tilde\sigma_l)} \\
& \stackrel{\phantom{\tn{SVD}}}{=}
\sum_{s_l}\, [M^{\sigma_l}]_{a_l,s_l} [\tilde M^{\tilde\sigma_l}]_{s_l,a_{l+1}}\,,
\end{split}\end{equation}
the (appropriately normalized) matrices $M=U$ and $\tilde M=SV$ define a MPS whose local Hilbert space dimension is reduced back to $d$. However, $M^{\sigma_l}$ and $M^{\sigma_{l+1}}$ are now next-nearest neighbors, and $e^{iHt}$ therefore contains longer-ranged interactions (the same holds for $e^{-iHt}$ which couples the $\tilde M$). Computing the time evolution $A(t)$ is thus just as easy (or hard) as it is to simulate dynamics at finite temperatures (note that now \textit{all sites} are physical ones). One can exploit $S^z$-conservation (or any other Abelian symmetry) by simply performing a spin-flip transformation $\tilde\uparrow\to\tilde\downarrow$, which then allows to straightforwardly employ the usual quantum numbers $m(\sigma)=\sigma$, $m(\tilde\sigma)=\tilde\sigma$. The numerical cost to carry out the time evolution $A(t)$ scales as $d^4\chi^3$. 

It seems instructive to illustrate the algorithmic simplicity of the second approach for two concrete examples: In order to calculate $S^z_l(t)$ or the kinetic energy $E_\tn{K}=-L[S^z_lS^z_{l+1}](t)/\Delta$ for a Heisenberg chain, one needs to time-evolve the states
\begin{equation}\label{eq:pura}\begin{split}
|\psi_{S^z}\rangle & = |\psi_{\mathbbm{1}}\rangle\otimes\cdots\otimes |\psi_{\mathbbm{1}}\rangle\otimes
|\psi_{S^z+1/2}\rangle \otimes |\psi_{\mathbbm{1}}\rangle \otimes\cdots\otimes |\psi_{\mathbbm{1}}\rangle\,, \\\
|\psi_{E_\tn{K}}\rangle & = |\psi_{\mathbbm{1}}\rangle\otimes\cdots\otimes |\psi_{\mathbbm{1}}\rangle\otimes
|\psi_{S^z+1/2}\rangle \otimes |\psi_{S^z+1/2}\rangle \otimes |\psi_{\mathbbm{1}}\rangle \otimes\cdots\otimes |\psi_{\mathbbm{1}}\rangle\,, \\\
|\psi_{\mathbbm{1}}\rangle & =  |\hspace*{-0.1cm}\uparrow\rangle\,|\hspace*{-0.0cm}\tilde\downarrow\rangle +  |\hspace*{-0.1cm}\downarrow\rangle\,|\hspace*{-0.0cm}\tilde\uparrow\rangle\,,~
|\psi_{S^z+1/2}\rangle = |\hspace*{-0.1cm}\uparrow\rangle\,|\hspace*{-0.0cm}\tilde\downarrow\rangle\,
\end{split}\end{equation}
under $e^{\pm iHt}$ which now contain purely next-nearest neighbor interactions between either only odd or even sites. This is almost completely equivalent to simulating a time evolution $e^{-iHt}|\psi_T\rangle$ at finite temperature and can thus be achieved directly with any code that can access $T>0$ dynamics.

While carrying out the time evolution scales as $\chi^3$ in both approaches, computing $\langle\psi| e^{iHt} A(t) e^{-iHt} |\psi\rangle$ requires an overlap calculation which scales as $d^2\chi^4$. Thus, the latter will eventually dominate the numerical effort and render it impossible to combine the Schr\"odinger and Heisenberg pictures. As already illustrated by Figures \ref{fig:neel} and \ref{fig:wavep} (whose data was obtained effortlessly), this is not the case in most practical applications. The reason for this pragmatic observation is two-fold: First, computing an overlap involves matrix multiplications, which are \textit{significantly} faster than carrying out a singular value decomposition (the bottleneck in the time evolution algorithms), even if one employs a SVD routine from highly optimized libraries such as Intel's Math Kernel Library. Moreover, the overlap calculation is highly parallelizable and scales much better with the number of low-level threads (typically 24 in our case) than the SVD for bond dimensions that occur in practice. Second, the additional factor of $\chi$ needs to be compared to the (typically) much faster growth of $\chi$ due to the buildup of entanglement: computing $S^z(t)$ in Figure \ref{fig:neel}(a) up to times $t\sim20$ within the Schr\"odinger picture would require a bond dimension of roughly $\chi\sim30000$ as opposed to $\chi\sim900$ if both pictures are combined.

\section{Analytic understanding of the finite-temperature disentangler}
\label{sec:hq}

It was shown in Refs.~\cite{drudepaper,dmrgpaper} that one can exploit the fact that purification is not unique to push finite-temperature simulations to larger times:
\begin{equation}\label{eq:corrt2}
C^{AB}_T(t) = \tn{Tr}\,\big[\rho_T A(t)B\big] =
\langle\psi_T|A(t)B |\psi_T\rangle = \langle\psi_T|U_Q^\dagger A(t)U_Q^{\phantom{\dagger}}B |\psi_T\rangle\,.
\end{equation}
For didactic reasons, we will now give the simplest possible (yet missed in Ref.~\cite{drudepaper}) analytic argument for why the particular choice $U_Q(t) = e^{i H_Q t}$ -- i.e., evolving $Q$ backwards in time with the physical Hamiltonian $H_Q$ acting on the auxiliary degrees of freedom -- succeeds in reducing the growth of entanglement (see also Ref.~\cite{trick2b}, which introduced a systematic way to further optimize $U_Q(t)$, as well as the discussion in the introduction of our paper). We present a straightforward way to determine whether or not the sign of any terms in $H_Q$ need to be flipped if for a given model Abelian symmetries are incorporated within the MPS numerics.

Let us begin with the Heisenberg chain defined in Eq.~(\ref{eq:h}) and try to analytically understand the time evolution of the state $|\psi_T\rangle$ which purifies the density matrix. Since $|\psi_T\rangle\sim e^{-H/T}|\psi_\infty\rangle$ and $H$ commutes with $H_Q$, it is sufficient to study $e^{-iHt}e^{iH_Qt}|\psi_\infty\rangle$. Starting from Ref.~\cite{dmrgrev}, we need to choose $|\psi_\infty\rangle$ as follows in order to exploit the conservation of the total (physical plus auxiliary) spin:
\begin{equation}
|\psi_\infty\rangle = \bigotimes_{l=1}^L |\psi_{\infty,l}\rangle\,,~|\psi_{\infty,l}\rangle= \frac{1}{\sqrt{2}} (|\hspace*{-0.1cm}\uparrow_l\downarrow_{l,Q}\rangle-|\hspace*{-0.1cm}\downarrow_l\uparrow_{l,Q}\rangle)\,,
\end{equation}
where we have combined physical and auxiliary degrees of freedom $\sigma_l$ and $\sigma_{l,Q}$ to a single site whose Hilbert space is spanned by $\{|\hspace*{-0.1cm}\uparrow\uparrow\rangle,|\hspace*{-0.1cm}\uparrow\downarrow\rangle,|\hspace*{-0.1cm}\downarrow\uparrow\rangle,|\hspace*{-0.1cm}\downarrow\downarrow\rangle \}$. Applying $H\otimes\mathbbm{1}$ then induces the following transitions between neighboring sites:
\begin{equation}\label{eq:trans1}
\begin{split}
\downarrow\uparrow |\uparrow\downarrow \;\;&\stackrel{(1/2)(S^+\otimes \mathbbm{1})\otimes (S^-\otimes \mathbbm{1})}{\longrightarrow}\;\;(1/2)\uparrow\uparrow |\downarrow\downarrow\\ 
\uparrow\downarrow |\downarrow\uparrow\;\;&\stackrel{(1/2)(S^-\otimes \mathbbm{1})\otimes (S^+\otimes \mathbbm{1})}{\longrightarrow}\;\;(1/2)\downarrow\downarrow|\uparrow\uparrow \\ 
x |y\;\;&\stackrel{(\Delta)(S^z\otimes \mathbbm{1})\otimes (S^z\otimes \mathbbm{1})}{\longrightarrow}\;\;{\rm sign_1}(x){\rm sign_1}(y) (\Delta/4)\;\;x|y \\
x |y\;\;&\stackrel{(b)(S^z\otimes \mathbbm{1})\otimes(\mathbbm{1}\otimes \mathbbm{1})}{\longrightarrow}\;\;(b/2){\rm sign_1}(x) \;\;x|y \,,
\end{split}
\end{equation}
where $x,y\in\{\uparrow\downarrow,\downarrow\uparrow\}$, ${\rm sign_1}(\uparrow\downarrow)=1 $, and ${\rm sign_1}(\downarrow\uparrow)=-1 $. The time evolution of $|\psi_\infty\rangle$ governed by $H\otimes\mathbbm{1}$ is thus non-trivial; entanglement starts to build up. However, applying $-\mathbbm{1}\otimes H_Q$ induces the same transitions but with opposite sign and hence undoes this damage:
\begin{equation}
\begin{split}
\downarrow\uparrow |\uparrow\downarrow \;\;&\stackrel{(-1/2)(\mathbbm{1}\otimes S^-)\otimes (\mathbbm{1}\otimes S^+)}{\longrightarrow}\;\;(-1/2)\downarrow\downarrow|\uparrow\uparrow \\ 
\uparrow\downarrow |\downarrow\uparrow\;\;&\stackrel{(-1/2)( \mathbbm{1}\otimes S^+)\otimes (\mathbbm{1}\otimes S^-)}{\longrightarrow}\;\;(-1/2)\uparrow\uparrow|\downarrow\downarrow \\ 
x |y\;\;&\stackrel{(-\Delta)(\mathbbm{1}\otimes S^z)\otimes (\mathbbm{1}\otimes S^z)}{\longrightarrow}\;\;[-{\rm sign_1}(x)][-{\rm sign_1}(y)] (-\Delta/4)\;\;x|y \\
x |y\;\;&\stackrel{(b)(\mathbbm{1}\otimes S^z)\otimes(\mathbbm{1}\otimes \mathbbm{1})}{\longrightarrow}\;\;(b/2)[-{\rm sign_1}(x)] \;\;x|y \,,
\end{split}
\end{equation}
\textit{if additionally the sign of the magnetic field $b$ is reversed in $H_Q$}. The total time evolution governed by $H\otimes\mathbbm{1}-\mathbbm{1}\otimes H_Q$ therefore becomes trivial, or put differently: While $|\psi_\infty\rangle$ is not an eigenstate of $H$ alone, it is an eigenstate of $H-H_Q$ (and the same holds for $|\psi_T\rangle$). The calculation of $A(t)|\psi_T\rangle$ is therefore only plagued by an entanglement building up around the region where $A$ acts (the physical reason being quasi-locality \cite{trick2b,trick2a}). This simple fact, which was missed in Ref.~\cite{drudepaper}, explains why the particular choice $U_Q(t)=e^{iH_Qt}$ reduces the growth of entanglement during calculations at finite temperature (see also Ref.~\cite{trick2b}). The Hubbard model defined in Eq.~(\ref{eq:h2}) can be analyzed analogously, and it turns out that if $|\psi_\infty\rangle$ is chosen such that one can exploit the conservation of both the total spin and the total charge, the signs of $\mu$ and $b$ need to be reversed in $H_Q$.

\begin{figure}[t]
\centering\includegraphics[width=0.48\linewidth,clip]{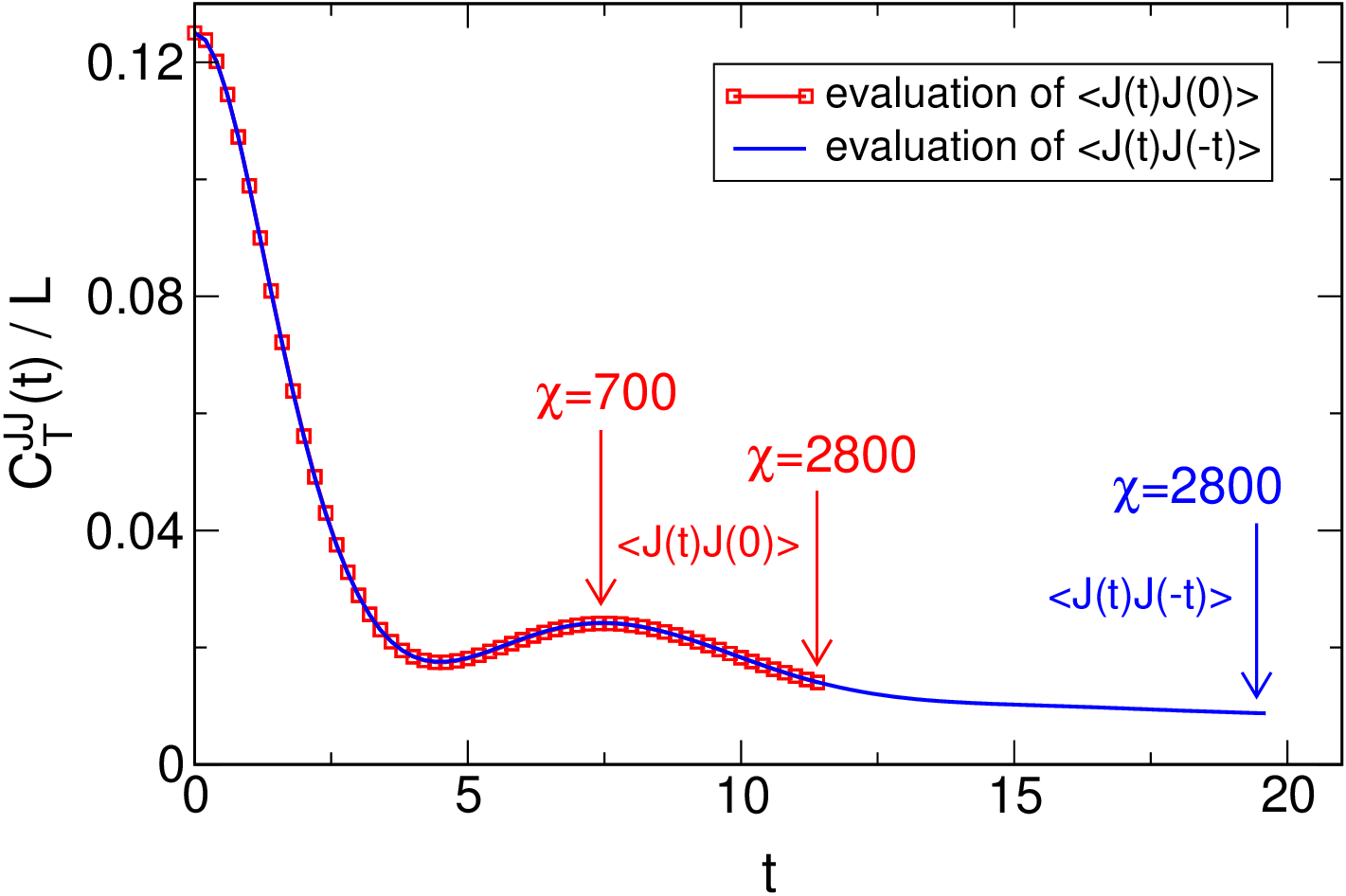}
\caption{(Color online) Global spin current correlation function of the isotropic XXZ chain ($\Delta=1$, $b=0$) at infinite temperature. If one `exploits time translation invariance' $C^{JJ}_T(2t)=\tn{Tr}[\rho_T J(t)J(-t)]$, one can access time scales twice as large at virtually no additional effort \cite{trick2b}. The calculation can be carried out efficiently within a MPS based DMRG code (see Sec.~\ref{sec:fac2}).}
\label{fig:jj}
\end{figure}

\section{A factor of two}
\label{sec:fac2}

To the best of our knowledge, it was overlooked for a long time (and only recently noticed in Ref.~\cite{trick2b} in an implicit way in the specific context of optimizing calculations at $T>0$) that one can simply `exploit time translation invariance' in equilibrium correlation functions to extend the range of simulations by a factor of two:
\begin{equation}\label{eq:trickC2}\begin{split}
C_\tn{gs}^{AB}(2t) & = \langle\tn{gs}| A(t) B(-t) | \tn{gs}\rangle\,, \\ C^{AB}_T(2t) & = \tn{Tr}\,\big[\rho_T A(t)B(-t)\big] =
\langle\psi_T|A(t)B(-t) |\psi_T\rangle\,.
\end{split}\end{equation}
At $T=0$ one needs to carry out two individual calculations for $e^{iHt}A|\tn{gs}\rangle$ as well as $e^{-iHt}B|\tn{gs}\rangle$, which can be done straightforwardly using MPS DMRG (at least if $A$ and $B$ are local operators; we will come back to this below). At $T>0$, one needs to perform a separate calculation of $e^{-iHt_n}Ae^{iHt_n}|\psi_T\rangle$ for every $t_n\in[0,t]$ (and likewise for $B$), which is possible \cite{drudepaper2} but costly. However, given the `insights' of the previous section that $|\psi_T\rangle$ is an eigenstate of $H-H_Q$, we can recast Eq.~(\ref{eq:trickC2}) as
\begin{equation}\label{eq:backt}
C^{AB}_T(2t) =
\langle\psi_T|e^{iHt} A e^{-iHt} e^{-iHt} B e^{-iHt} |\psi_T\rangle =
\langle\psi_T|A e^{-iHt+iH_Qt} e^{-iHt+iH_Qt} B |\psi_T\rangle\,,
\end{equation}
and are left with the significantly simpler task to calculate $e^{-iHt+iH_Qt}A|\psi_T\rangle$ (as well as $e^{iHt-iH_Qt}B|\psi_T\rangle$) via a single DMRG simulation up to the time $t$. Note that the `disentangler' $U_Q(t)$ is automatically included in Eq.~(\ref{eq:backt}).

In case that $A$ is given by a product of local operators (e.g., $S^z_l$ or $S^+_lS^-_{l+1}$), it can be applied straightforwardly to a MPS representation of $|\tn{gs}\rangle$ or $|\psi_T\rangle$ without increasing its dimension $\chi$. If $A$ contains an arbitrary sum of products of local operators $a_n$, one could simply carry out a separate DMRG calculation for each $e^{iHt}a_n|\tn{gs}\rangle$ or $e^{-iHt+iH_Qt}a_n|\psi_T\rangle$. This is possible in principle \cite{drudepaper2} but typically increases the computational effort by a factor of the order of the system size. It can be avoided trivially in case that $A$ can be expressed in terms of a MPO with a small bond dimension. We will now discuss one instructive example for didactic purposes.

Let us assume that we want to calculate the autocorrelator $C^{JJ}_T(2t)$ of the global spin current $J=\sum_l j_l$, $j_l=(S^+_lS^-_{l+1}-S^-_lS^+_{l+1})/2i$ for the XXZ chain defined in Eq.~(\ref{eq:h}). Spatial translation invariance stipulates $C^{JJ}_T(2t)=LC^{j_{L/2}J}_T(2t)$, and in absence of a magnetic field ($b=0$) spin flip symmetries yields $C^{j_{L/2}J}_T(2t)=2C^{j_{L/2}J_\uparrow}_T(2t)$, where $J=J_\uparrow+J_\uparrow^\dagger$. Instead of computing each term in $e^{-iHt+iH_Qt}J_\uparrow|\psi_T\rangle$ individually \cite{drudepaper2}, one can express $J_\uparrow$ in terms of a matrix product operator $J_\uparrow = \prod_{l=1}^{L} J_\uparrow^{[l]}$ with
\begin{equation}
J_\uparrow^{[1]}=
\begin{pmatrix}
0 & \frac{S^+_1}{2i} & 0       
\end{pmatrix},~J_\uparrow^{[l=2\ldots L-1]}=
\begin{pmatrix}
1 & 0 & 0 \\ S^-_l & 0 & 0 \\ 0 & \frac{S^+_l}{2i} & 0       
\end{pmatrix},~J_\uparrow^{[L]}=
\begin{pmatrix}
1 \\ S^-_L \\ 0       
\end{pmatrix}\,.
\end{equation}
Applying $J_\uparrow\otimes\mathbbm{1}_Q$ to $|\psi_T\rangle$ yields a new MPS (with a bond dimension increased by a factor of 3), which can then be time evolved via $e^{-iHt+iH_Qt}$. This altogether illustrates that Eq.~(\ref{eq:trickC2}) can be incorporated readily. An example for how this `trick' allows to reach larger time scales is shown in Figure \ref{fig:jj}.

\section{The beauty of python}

In this section, we illustrate the numerical simplicity of the core DMRG algorithms (thoroughly described in Ref.~\cite{dmrgrev}) if implemented within the python programming language. We hope to advocate the method to colleagues new to its realm and try to stimulate the development of new DMRG codes.

Let us assume that we want to time evolve a given MPS whose matrices $M^{\sigma_l}$ are expressed in the `$\Lambda-\Gamma$ form' 
\begin{equation}
M^{\sigma_l}_{a_la_{l+1}} =  \Lambda_{a_l}^l\Gamma_{a_la_{l+1}}^{\sigma_l}~.
\end{equation}
As usual, the local Hilbert space and (position-dependent) bond dimensions are denoted by $d$ and $\chi_l$, respectively. After a Trotter decomposition of $e^{-iHt}$, the key task is to apply local operators $O(\sigma_l,\sigma_{l+1};\tilde\sigma_l,\tilde\sigma_{l+1}')$ to Eq.~(\ref{eq:mps}). The first step (labeled `step 1' in the python code below) is to form the three-site wave function
\begin{equation}
\Psi^{\sigma_0\sigma_1} = \Lambda^0\Gamma^{\sigma_0}\Lambda^1\Gamma^{\sigma_1}\Lambda^2\,,
\end{equation}
where we have randomly set $l=0$ to keep the notation simple. Next (step 2), we apply $O(\sigma_0,\sigma_{1};\tilde\sigma_0,\tilde\sigma_{1})$,
\begin{equation}
\Phi_{a_0,a_2}^{\sigma_0\sigma_1} = \sum_{\tilde\sigma_0\tilde\sigma_1}O(\sigma_0,\sigma_{1};\tilde\sigma_0,\tilde\sigma_{1})\Psi^{\tilde\sigma_0\tilde\sigma_1}_{a_0,a_2}\,,
\end{equation}
and carry out a singular value decomposition (SVD) of the appropriately reshaped tensor $\Phi_{\sigma_0a_0,\sigma_1a_2}$ (step 3):
\begin{equation}
\Phi_{\sigma_0a_0,\sigma_1a_2} = \sum_{a_1} U_{\sigma_0a_0,a_1}S_{a_1} V_{s_1,\sigma_1a_2}\,.
\end{equation}
The updated matrices $\tilde\Gamma^{\sigma_0}$, $\tilde\Lambda^1$, and $\tilde\Gamma^{\sigma_1}$ are then obtained as (step 4)
\begin{equation}
\tilde\Gamma^{\sigma_0}_{a_0,a_1} = U_{\sigma_0a_0,a_1}/\Lambda^0_{a_0} \,,~
\tilde\Lambda^1_{a_1} = S_{a_1}\,,~ 
\tilde\Gamma^{\sigma_1}_{a_1,a_2} = V_{a_1,\sigma_1a_2}/\Lambda^2_{a_2} \,,
\end{equation}
where we ignore the subtlety of a numerical division by (potentially) small singular values \cite{hastings}. The bond dimension $\chi_1$ increased by a factor of $d$; it is usually truncated down to a given $\chi_{1,\tn{max}}$, and the associated error is controlled by the discarded weight
\begin{equation}
\tn{discarded} = \sum_{a_1=\chi_{1,\tn{max}}+1}^{d\chi_1} (S_{a_1})^2\,.
\end{equation}
The above steps can be implemented straightforwardly within python:

\begin{lstlisting}[language=Python,basicstyle=\footnotesize]
import numpy as np

def bond(l0, G0, l1, G1, l2, d, chi1_max, O):

  # initial bond dimensions
  chi0=len(l0); chi2=len(l2)
    
  # step 1: form 3-site tensor
  Psi = np.tensordot( np.diag(l0), G0         , axes=(1,1) )
  Psi = np.tensordot( Psi        , np.diag(l1), axes=(2,0) )
  Psi = np.tensordot( Psi        , G1         , axes=(2,1) )
  Psi = np.tensordot( Psi        , np.diag(l2), axes=(3,0) )
    
  # step 2: apply local time evolution operator
  Phi = np.tensordot( Psi, O, axes=([1,2],[0,1]) )
  Phi = np.transpose( Phi, (2,0,3,1)     )
  Phi = np.reshape  ( Phi, (d*chi0,d*chi2) )
    
  # step 3: singular value decomposition
  U, S, V = np.linalg.svd(Phi,0)
            
  # step 4: truncate and reshape
  chi1 = min( len(S), chi1_max )
  
  discarded = np.sum(S[chi1:]**2) / np.sum(S**2)
  tilde_l1 = S[0:chi1] / np.sqrt(np.sum(S[0:chi1]**2))
    
  U        = np.reshape  ( U[:,0:chi1], (d,chi0,chi1) )
  U        = np.tensordot( U, np.diag(1.0/l0), axes=(1,0) )
  tilde_G0 = np.transpose( U, (0,2,1) )
    
  V        = np.reshape  ( (V.T)[:,0:chi1], (d,chi2,chi1)  )
  tilde_G1 = np.tensordot(  V, np.diag(1.0/l2), axes=(1,0) )
        
  return tilde_G0, tilde_l1, tilde_G1, discarded
\end{lstlisting}

where $\Lambda^l$ and $\Gamma^l$ are arrays of size $\chi_l$ and $(d,\chi_l,\chi_{l+1})$, respectively. The by far most time consuming parts of this algorithm are matrix multiplications (`tensordot') as well as the singular value decomposition, which are elegantly wrapped to lapack routines through the `numpy' package. The numerical overhead due to the inefficiency of python is almost completely negligible. Significant speedups can be achieved 1) by compiling numpy using an efficient and parallelized implementation of lapack (e.g., Intel's Math Kernel Library), and 2) by carrying out the time evolution on different bonds in parallel via high-level parallelization in python. Generalizing the above routine to a next-nearest neighbor interaction is extremely straightforward. Abelian symmetries can be incorporated readily and speed up calculations by a factor of $\sim10$ for a bond dimension of $\chi=1000$ (exploiting continuous symmetries, however, is a challenging task \cite{sym1,sym2,sym3}).

\section{Summary}

We discussed a few simple tricks to extend the range of time-dependent DMRG simulations. In particular, we illustrated 1) how to `combine' the Schr\"odinger- and Heisenberg picture in the evaluation of $\langle\psi|A(t)|\psi\rangle$ as well as for similar quench dynamics at finite temperatures, 2) how to compute $A(t)$ within a MPS based DMRG code, 3) how to efficiently exploit time translation invariance in equilibrium correlation functions, and 4) how to analytically understand why a recently-introduced disentangler succeeds in reducing the entanglement growth of calculations at $T>0$. Aiming at beginners, we presented the most important aspects of how to implement these tricks in practice.

\vspace*{1ex}

\emph{Acknowledgments} --- We are grateful to Thomas Barthel and Volker Meden for useful suggestions. Support by the Nanostructured Thermoelectrics program of LBNL (CK) is acknowledged.

\vspace*{3ex}


\end{document}